\documentclass[a4,12pt]{article}

\usepackage{graphicx}

\begin{document}

\vspace{0.2cm}

\begin{center}
{\large\bf Flavor Mixing Democracy and Minimal CP Violation}
\end{center}

\vspace{0.1cm}

\begin{center}
{\bf Jean-Marc Gerard}$^\dagger$\footnote{E-mail: jean-marc.gerard@uclouvain.be},
{\bf Zhi-zhong Xing}$^\ddagger$\footnote{E-mail: xingzz@ihep.ac.cn} \\
$^\dagger${Centre for Cosmology, Particle Physics and Phenomenology (CP3), \\
Universite Catholique de Louvain, B-1348, Louvain-la-Neuve,
Belgium} \\
$^\ddagger${Institute of High Energy Physics, Chinese Academy of
Sciences, Beijing 100049, China}
\end{center}

\vspace{1.5cm}

\begin{abstract}
We point out that there is a unique parametrization of quark flavor
mixing in which every angle is close to the Cabibbo angle
$\theta^{}_{\rm C} \simeq 13^\circ$ with the CP-violating phase
$\phi^{}_{q}$ around $1^\circ$, implying that they might all be
related to the strong hierarchy among quark masses. Applying the
same parametrization to lepton flavor mixing, we find that all three
mixing angles are comparably large (around $\pi/4$) and the Dirac
CP-violating phase $\phi^{}_{l}$ is also minimal as compared with
its values in the other eight possible parametrizations. In this
spirit, we propose a simple neutrino mixing ansatz which is
equivalent to the tri-bimaximal flavor mixing pattern in the
$\phi^{}_{l} \to 0$ limit and predicts $\sin\theta^{}_{13} =
1/\sqrt{2} \sin (\phi^{}_{l}/2)$ for reactor antineutrino
oscillations. Hence the Jarlskog invariant of leptonic CP violation
$J^{}_{l} = (\sin\phi^{}_{l})/12$ can reach a few percent if
$\theta^{}_{13}$ lies in the range $7^\circ \leq \theta^{}_{13} \leq
10^\circ$.
\end{abstract}


\def\thefootnote{\arabic{footnote}}
\setcounter{footnote}{0}

\newpage

\section{Introduction}

Within the standard electroweak model, the origin of CP violation is
attributed to an irremovable phase of the $3\times 3$
Cabibbo-Kobayashi-Maskawa (CKM) quark flavor mixing matrix
\cite{CKM} in the charged-current interactions:
\begin{eqnarray}
-{\cal L}^{q}_{\rm cc} = \frac{g}{\sqrt{2}} \ \overline{
\left(\matrix{u & c & t} \right)^{}_{\rm L}} \ \gamma^\mu
\left( \matrix{ V^{}_{ud} & V^{}_{us} & V^{}_{ub} \cr
V^{}_{cd} & V^{}_{cs} & V^{}_{cb} \cr
V^{}_{td} & V^{}_{ts} & V^{}_{tb} \cr} \right)
\left(\matrix{ d \cr s \cr b \cr} \right)^{}_{\rm L} W^+_\mu \
+ \ {\rm h.c.} \; .
\end{eqnarray}
The size of this nontrivial CP-violating phase depends on the
explicit parametrization of the CKM matrix $V$. One may in general
describe $V$ in terms of three rotation angles and one CP-violating
phase, and arrive at nine topologically different parametrizations
\cite{FX98}. If $V$ takes the Cabibbo flavor mixing pattern
\cite{Cabibbo78}
\begin{eqnarray}
V^{}_{\rm C} = \frac{1}{\sqrt{3}} \left( \matrix{ 1 & 1 & 1 \cr 1 &
\omega & \omega^2 \cr 1 & \omega^2 & \omega \cr} \right) \; ,
\end{eqnarray}
where $\omega = e^{i2\pi/3}$ is the complex cube-root of unity
(i.e., $\omega^3 =1$), then one can immediately find that the
CP-violating phases in all the nine parametrizations are exactly
$\pi/2$. Hence $V^{}_{\rm C}$ characterizes the case of ``maximal CP
violation" in a parametrization-independent way, although it is not
a realistic quark flavor mixing matrix. Among the nine
parametrizations of $V$ listed in Ref. \cite{FX98}, the one
advocated by the Particle Data Group \cite{PDG} is most popular and
its CP-violating phase is about $65^\circ$. The idea of a
``geometrical T violation" has been suggested in Ref. \cite{Branco}
to explain such a CP-violating phase around $\pi/3$. In comparison,
the CP-violating phase is about $90^\circ$ in the parametrization
recommended in Ref. \cite{FX97} or in the original Kobayashi-Maskawa
representation \cite{He}. Accordingly, the concept of ``maximal CP
violation" has sometimes been used to refer to a quark flavor mixing
scenario in which the CP-violating phase equals $\pi/2$ for given
values of the mixing angles \cite{FX95,MCP,Xing09,Others}.

Of course, the value of the CP-violating phase is correlated with
the values of the mixing angles in a given parametrization of $V$.
Indeed, the parametrization itself depends on the chosen flavor
basis and only the moduli of the matrix elements $V^{}_{i j}$ are
completely basis-independent. Although all the parametrizations of
$V$ are mathematically equivalent, one of them might be
phenomenologically more interesting in the sense that it might
either make the underlying physics of quark mass generation and CP
violation more transparent or lead to more straightforward and
simpler relations between the fundamental flavor mixing parameters
and the corresponding observable quantities. It is therefore
meaningful to examine different parametrizations of the CKM matrix
$V$ and single out the one which is not only phenomenologically
useful but also allows us to have a new insight into the flavor
puzzles and possible solutions to them.

In this paper we pose such a question: is it possible to ascribe
small CP-violating effects in the quark sector to a strongly
suppressed CP-violating phase in the CKM matrix $V$ in which all
three mixing angles are comparably sizable? The answer to this
question is actually affirmative as already observed in Ref.
\cite{G08}, and the details of such a nontrivial description of
quark flavor mixing and CP violation will be elaborated in section
2. We show that the CP-violating phase $\phi^{}_q$ is only about
$1^\circ$, while every quark mixing angle is close to the Cabibbo
angle $\theta^{}_{\rm C} \simeq 13^\circ$ in this unique
parametrization of $V$, implying that they might all have something
to do with the strong hierarchy of quark masses. We argue that this
particular representation reveals an approximate flavor mixing
democracy and ``minimal CP violation". It also provides a simple
description of the structure of the matrix $V$, which is almost
symmetric in modulus about its $V^{}_{ud}$-$V^{}_{cs}$-$V^{}_{tb}$
axis.

Applying the same parametrization to the lepton flavor mixing, we
find that all three angles are comparably large (around $\pi/4$) and
the Dirac CP-violating phase $\phi^{}_l$ is also minimal as compared
with its values in the other eight possible parametrizations. We
start from this observation to propose a simple and testable
neutrino mixing ansatz which is equal to the well-known
tri-bimaximal flavor mixing pattern \cite{TB} in the $\phi^{}_l \to
0$ limit. It predicts $\sin\theta^{}_{13} = 1/\sqrt{2} \sin
\left(\phi^{}_l/2\right)$ for reactor antineutrino oscillations, and
its two larger mixing angles are consistent with solar and
atmospheric neutrino oscillations. The Jarlskog invariant for
leptonic CP violation turns out to be $J^{}_l =
\left(\sin\phi^{}_l\right)/12$, which can reach a few percent if
$\theta^{}_{13}$ lies in the range $7^\circ \leq \theta^{}_{13} \leq
10^\circ$.

\section{Quark flavor mixing}

The parametrization of the CKM matrix $V$, which assures an
approximate flavor mixing democracy and nearly minimal CP violation
in the quark sector, takes the form
\begin{eqnarray}
V \hspace{-0.2cm} & = & \hspace{-0.2cm}
\left( \matrix{c^{}_y & 0 & s^{}_y \cr
0 & 1 & 0 \cr -s^{}_y & 0 & c^{}_y \cr} \right)
\left( \matrix{ c^{}_x & s^{}_x & 0 \cr
-s^{}_x & c^{}_x & 0 \cr
0 & 0 & e^{-i\phi^{}_q} \cr} \right)
\left( \matrix{c^{}_z & 0 & -s^{}_z \cr
0 & 1 & 0 \cr s^{}_z & 0 & c^{}_z \cr} \right)
\nonumber \\
\hspace{-0.2cm} & = & \hspace{-0.2cm}
\left( \matrix{c^{}_x c^{}_y c^{}_z + s^{}_y s^{}_z e^{-i\phi^{}_q}
& s^{}_x c^{}_y & -c^{}_x c^{}_y s^{}_z + s^{}_y c^{}_z e^{-i\phi^{}_q} \cr
-s^{}_x c^{}_z & c^{}_x & s^{}_x s^{}_z \cr
-c^{}_x s^{}_y c^{}_z + c^{}_y s^{}_z e^{-i\phi^{}_q} & -s^{}_x s^{}_y
& c^{}_x s^{}_y s^{}_z + c^{}_y c^{}_z e^{-i\phi^{}_q} \cr} \right) \; ,
\end{eqnarray}
where $c^{}_x \equiv \cos\theta^{}_x$ and $s^{}_x \equiv
\sin\theta^{}_x$, and so on. Without loss of generality, we arrange
the mixing angles to lie in the first quadrant but allow the
CP-violating phase $\phi^{}_q$ to vary between zero and $2\pi$.
Comparing Eq. (1) with Eq. (3), we immediately arrive at the
relation $\cos\theta^{}_x = |V^{}_{cs}|$ together with
\begin{eqnarray}
\tan\theta^{}_y \hspace{-0.2cm} & = & \hspace{-0.2cm}
\left|\frac{V^{}_{ts}}{V^{}_{us}}\right| \; , \nonumber \\
\tan\theta^{}_z \hspace{-0.2cm} & = & \hspace{-0.2cm}
\left|\frac{V^{}_{cb}}{V^{}_{cd}}\right| \; .
\end{eqnarray}
In this parametrization the off-diagonal asymmetries of $V$ in modulus
\cite{Xing95} are given as
\begin{eqnarray}
\Delta^{q}_{\rm L} \hspace{-0.2cm} & \equiv & \hspace{-0.2cm}
|V^{}_{us}|^2 - |V^{}_{cd}|^2 =
|V^{}_{cb}|^2 - |V^{}_{ts}|^2 = |V^{}_{td}|^2 - |V^{}_{ub}|^2 =
s^2_x \left( s^2_z - s^2_y \right) \; ,
\nonumber \\
\Delta^{q}_{\rm R} \hspace{-0.2cm} & \equiv & \hspace{-0.2cm}
|V^{}_{us}|^2 - |V^{}_{cb}|^2 =
|V^{}_{cd}|^2 - |V^{}_{ts}|^2 = |V^{}_{tb}|^2 - |V^{}_{ud}|^2 =
s^2_x \left( c^2_y - s^2_z \right) \; .
\end{eqnarray}
Note that the other eight parametrizations listed in Ref.
\cite{FX98} are unable to express $\Delta^{q}_{\rm L}$ and
$\Delta^{q}_{\rm R}$ in such a simple way. Furthermore, the Jarlskog
invariant for CP violation \cite{J} reads
\begin{eqnarray}
~~~~~ J^{}_q = {\rm Im}\left(V^{}_{ud} V^{}_{cs} V^*_{us}
V^*_{cd}\right) = {\rm Im}\left(V^{}_{us} V^{}_{cb} V^*_{ub}
V^*_{cs}\right) = \cdots = c^{}_x s^2_x c^{}_y s^{}_y c^{}_z s^{}_z
\sin\phi^{}_q \; .
\end{eqnarray}
We observe that choosing $|V^{}_{cs}|$, $\Delta^{q}_{\rm L}$,
$\Delta^{q}_{\rm R}$ and $J^{}_q$ as four independent parameters to
describe the CKM matrix $V$ is also an interesting possibility,
because they determine the geometric structure of $V$ and its CP
violation in a straightforward and rephasing-invariant manner.

To see the point that $\theta^{}_x$, $\theta^{}_y$ and $\theta^{}_z$
are comparable in magnitude, let us express them in terms of the
well-known Wolfenstein parameters \cite{Wol}. Up to the accuracy of
${\cal O}(\lambda^6)$, the Wolfenstein-like expansion of the CKM
matrix $V$ \cite{KM} is given as
\begin{eqnarray}
V \simeq \left( \matrix{ 1 - \frac{1}{2} \lambda^2 - \frac{1}{8}
\lambda^4 & \lambda & A\lambda^3 \left(\rho - i \eta\right) \cr
-\lambda \left[ 1 - A^2 \lambda^4 \left(\frac{1}{2} - \rho\right) +
i A^2 \lambda^4 \eta \right] & 1 - \frac{1}{2} \lambda^2 -
\frac{1}{8} \left(1 + 4A^2 \right) \lambda^4 & A\lambda^2 \cr A
\lambda^3 \left[1 - \left(1 - \frac{1}{2} \lambda^2 \right)
\left(\rho + i\eta \right) \right] & -A\lambda^2 \left[ 1 -
\lambda^2 \left( \frac{1}{2} - \rho\right) + i\lambda^2 \eta \right]
& 1 - \frac{1}{2} A^2 \lambda^4 \cr} \right) \; ,
\end{eqnarray}
where $\lambda = 0.2253 \pm 0.0007$, $A = 0.808^{+0.022}_{-0.015}$,
$\rho = 0.135^{+0.023}_{-0.014}$ and $\eta = 0.350\pm 0.013$
extracted from a global fit of current experimental data on flavor
mixing and CP violation in the quark sector \cite{PDG}. Comparing
Eq. (7) with Eq. (3), we arrive at the approximate relations
\begin{eqnarray}
\tan\theta^{}_x \hspace{-0.2cm} & \simeq & \hspace{-0.2cm} \lambda
\left[ 1 + \frac{1}{2} \left( 1 + A^2 \right) \lambda^2 \right] \; ,
\nonumber \\
\tan\theta^{}_y \hspace{-0.2cm} & \simeq & \hspace{-0.2cm}
A \lambda \left[ 1 - \frac{1}{2} \left( 1 - 2\rho \right) \lambda^2
\right] \; ,
\nonumber \\
\tan\theta^{}_z \hspace{-0.2cm} & \simeq & \hspace{-0.2cm}
A \lambda \; ,
\nonumber \\
\sin\phi^{}_q \hspace{-0.2cm} & \simeq & \hspace{-0.2cm}
\lambda^2 \eta \left[ 1 + \frac{1}{2}
\left(1 + 2A^2 - 2\rho \right) \lambda^2 \right] \; ,
\end{eqnarray}
which hold up to the accuracy of ${\cal O}(\lambda^5)$. Therefore,
we obtain
\begin{eqnarray}
~~~~ \theta^{}_x \simeq 13.2^\circ \; , ~~~~
\theta^{}_y \simeq 10.1^\circ \; , ~~~~
\theta^{}_z \simeq 10.3^\circ \; , ~~~~
\phi^{}_q \simeq 1.1^\circ \; .
\end{eqnarray}
We see that the small difference between
$\theta^{}_y$ and $\theta^{}_z$ signifies a slight off-diagonal
asymmetry of the CKM matrix $V$ in modulus about its
$V^{}_{ud}$-$V^{}_{cs}$-$V^{}_{tb}$ axis. Note that this tiny
asymmetry is quite
stable against the renormalization-group-equation (RGE)
running effects from the electroweak scale to a superhigh-energy
scale or vice versa. Indeed, only the Wolfenstein parameter
$A$ is sensitive to the RGE evolution \cite{RRR} so that
$\theta^{}_y$ and $\theta^{}_z$ run in the same way even at
the two-loop level
\footnote{We thank H. Zhang for confirming this point using the
two-loop RGEs of gauge and Yukawa couplings.}.
In contrast, $\theta^{}_x$ and $\phi^{}_q$ are almost insensitive
to the RGE running effects. The striking fact that
the CP-violating phase $\phi^{}_q$ is especially small in this
parametrization was first emphasized in Ref. \cite{G08}.
Indeed, the other eight parametrizations listed in Ref. \cite{FX98}
all require $\phi^{}_q \geq 60^\circ$. Moreover, the values
\begin{eqnarray}
~~~~ \Delta^{q}_{\rm L} \simeq 6.3 \cdot 10^{-5} \; , ~~~
\Delta^{q}_{\rm R} \simeq 4.9 \cdot 10^{-2} \; , ~~~ J^{}_q \simeq
3.0 \cdot 10^{-5} \;
\end{eqnarray}
indicate that $\theta^{}_y = \theta^{}_z$ and
$\phi^{}_q=0$ might be two good leading-order approximations from
the point of view of model building. In these two limits the CKM
matrix $V$ is real and symmetric in modulus. Consequently,
the small off-diagonal asymmetry
and the small CP-violating phase of $V$ might come from
some complex perturbations at the level of quark mass
matrices.

Why may $\phi^{}_q \sim \lambda^2$ coexist with $\theta^{}_x \sim
\theta^{}_y \sim \theta^{}_z \sim \lambda$? The reason is simply
that $V^{}_{ub}$ is the smallest CKM matrix element and only a small
$\phi^{}_q$ guarantees a significant cancellation in
$V^{}_{ub} = -c^{}_x c^{}_y s^{}_z + s^{}_y c^{}_z e^{-i\phi^{}_q}$ to
make $|V^{}_{ub}| \sim {\cal O}(\lambda^4)$ hold
\footnote{Because of $A \simeq 0.808$, $\rho \simeq 0.135$ and $\eta
\simeq 0.350$, the true order of $|V^{}_{ub}|$ is $\lambda^4$
instead of $\lambda^3$. Following the original spirit of the
Wolfenstein parametrization \cite{Wol}, one may consider to take
$V^{}_{ub} = A\lambda^4 \left(\hat{\rho} - i \hat{\eta}\right)$ by
redefining two ${\cal O}(1)$ parameters $\hat{\rho} = \rho/\lambda
\simeq 0.599$ and $\hat{\eta} = \eta/\lambda \simeq 1.553$.}.
The point that $V^{}_{ub}$ strongly depends on $\phi^{}_q$ motivates
us to propose a phenomenological ansatz for quark flavor mixing in
which $V^{}_{ub} \to 0$ holds in the $\phi^{}_q \to 0$ limit. In
this case we find that the condition $\tan\theta^{}_y =
\tan\theta^{}_z \cos\theta^{}_x$ must be fulfilled and the CKM matrix
reads
\begin{eqnarray}
V^{}_0 = \left( \matrix{s^{}_y /s^{}_z & s^{}_x c^{}_y & 0 \cr
-s^{}_x c^{}_z & c^{}_x & s^{}_x s^{}_z \cr s^2_x c^{}_y s^{}_z &
-s^{}_x s^{}_y & c^{}_z / c^{}_y \cr} \right) \; .
\end{eqnarray}
Of course, $V^{}_0$ can approximately describe the observed
moduli of the nine CKM matrix elements. The relation
$\tan\theta^{}_y = \tan\theta^{}_z \cos\theta^{}_x$ implies
that $\theta^{}_z$ must be slightly larger than $\theta^{}_y$,
and thus it has no conflict with the numerical results obtained
in Eq. (9). Now the CP-violating phase $\phi^{}_q$
is switched on and $V^{}_0$ is changed to
\begin{eqnarray}
V = \left( \matrix{ \left( c^2_z + s^{2}_z e^{-i\phi^{}_q}
\right) s^{}_y /s^{}_z & s^{}_x c^{}_y &
-s^{}_y c^{}_z \left( 1- e^{-i\phi^{}_q} \right) \cr
-s^{}_x c^{}_z & c^{}_x & s^{}_x s^{}_z \cr
-c^{}_y s^{}_z \left( c^2_x - e^{-i\phi^{}_q} \right)
& -s^{}_x s^{}_y
& \left( s^2_y + c^2_y e^{-i\phi^{}_q} \right) c^{}_z/c^{}_y
\cr} \right) \; ,
\end{eqnarray}
which predicts $|V^{}_{ub}| = 2s^{}_y c^{}_z \sin(\phi^{}_q/2)
\simeq s^{}_y c^{}_z \sin\phi^{}_q$ for very small $\phi^{}_q$.
Comparing Eq. (12) with Eq. (7), we arrive at $\tan\theta^{}_x
\simeq \lambda$, $\tan\theta^{}_y \simeq \tan\theta^{}_z
\simeq A\lambda$ and $\sin\phi^{}_q \simeq \lambda^2
\sqrt{\rho^2 + \eta^2}$ in the leading-order approximation.
We conclude that this ansatz is essentially valid, and
it provides us with a good lesson for dealing with
lepton flavor mixing in section 3.

It has long been speculated that the small quark flavor mixing
angles might be directly related to the strong quark mass
hierarchies \cite{Cabibbo,Fritzsch}, in particular when the quark
mass matrices possess a few texture zeros which can naturally
originate from a certain flavor symmetry \cite{FN}. In this sense
it is also interesting for us to consider possibly simple and
instructive relations between quark mass ratios ($m^{}_u/m^{}_c$,
$m^{}_c/m^{}_t$, $m^{}_d/m^{}_s$ and $m^{}_s/m^{}_b$) and flavor
mixing parameters ($\theta^{}_x$, $\theta^{}_y$, $\theta^{}_z$ and
$\phi^{}_q$) in the parametrization of $V$ under discussion. In view
of the values for the quark masses renormalized at the electroweak
scale \cite{XZZ}, we make the naive conjectures
\begin{eqnarray}
\sin\theta^{}_x \hspace{-0.2cm} & \simeq & \hspace{-0.2cm}
\sqrt{\frac{m^{}_d}{m^{}_s} + \frac{m^{}_u}{m^{}_c}} \; ,
\nonumber \\
\sin\theta^{}_y \hspace{-0.2cm} & \simeq & \hspace{-0.2cm}
\sin\theta^{}_z \simeq \sqrt{\frac{m^{}_d}{m^{}_s}} -
\sqrt{\frac{m^{}_u}{m^{}_c}} \; ,
\nonumber \\
\sin\phi^{}_q \hspace{-0.2cm} & \simeq & \hspace{-0.2cm}
\frac{m^{}_s}{m^{}_b} \; .
\end{eqnarray}
Of course, these approximate relations are only valid at the
electroweak scale, and whether they can easily be derived from a
realistic model of quark mass matrices remains an open question. But
a possible correlation between the smallness of the CP-violating
phase and the smallness of quark mass ratios (e.g., $\sin\phi^{}_q
\simeq m^{}_s/m^{}_b$ as first conjectured in Ref. \cite{G08}) is
certainly interesting and suggestive, because it might imply a
common origin for the quark mass spectrum, flavor mixing and CP
violation. We hope that such a phenomenological observation based
on our particular parametrization in Eq. (3) may be useful to infer the
presence of an underlying flavor symmetry from the experimental data
in the near future.

\section{Lepton flavor mixing}

We proceed to consider the $3\times 3$
Maki-Nakagawa-Sakata-Pontecorvo (MNSP) lepton flavor
mixing matrix \cite{MNS} in the weak charged-current interactions:
\begin{eqnarray}
-{\cal L}^{l}_{\rm cc} = \frac{g}{\sqrt{2}} \ \overline{
\left(\matrix{e & \mu & \tau} \right)^{}_{\rm L}} \ \gamma^\mu
\left( \matrix{ U^{}_{e1} & U^{}_{e2} & U^{}_{e3} \cr
U^{}_{\mu 1} & U^{}_{\mu 2} & U^{}_{\mu 3} \cr
U^{}_{\tau 1} & U^{}_{\tau 2} & U^{}_{\tau 3} \cr} \right)
\left(\matrix{ \nu^{}_1 \cr \nu^{}_2 \cr \nu^{}_3 \cr} \right)^{}_{\rm L}
W^-_\mu \ + \ {\rm h.c.} \; .
\end{eqnarray}
The MNSP matrix $U$ can be parametrized in the same way as in Eq. (3):
\begin{eqnarray}
U \hspace{-0.2cm} & = & \hspace{-0.2cm} \left( \matrix{c^{}_b & 0 &
s^{}_b \cr 0 & 1 & 0 \cr -s^{}_b & 0 & c^{}_b \cr} \right) \left(
\matrix{ c^{}_a & s^{}_a & 0 \cr -s^{}_a & c^{}_a & 0 \cr 0 & 0 &
e^{-i\phi^{}_l} \cr} \right) \left( \matrix{c^{}_c & 0 & -s^{}_c \cr
0 & 1 & 0 \cr s^{}_c & 0 & c^{}_c \cr} \right) P^{}_\nu
\nonumber \\
\hspace{-0.2cm} & = & \hspace{-0.2cm} \left( \matrix{c^{}_a c^{}_b
c^{}_c + s^{}_b s^{}_c e^{-i\phi^{}_l} & s^{}_a c^{}_b & -c^{}_a
c^{}_b s^{}_c + s^{}_b c^{}_c e^{-i\phi^{}_l} \cr -s^{}_a c^{}_c &
c^{}_a & s^{}_a s^{}_c \cr -c^{}_a s^{}_b c^{}_c + c^{}_b s^{}_c
e^{-i\phi^{}_l} & -s^{}_a s^{}_b & c^{}_a s^{}_b s^{}_c + c^{}_b
c^{}_c e^{-i\phi^{}_l} \cr} \right) P^{}_\nu \; ,
\end{eqnarray}
where $P^{}_\nu = {\rm Diag}\{e^{i\rho}, e^{i\sigma}, 1\}$ denotes
an irremovable phase matrix if the massive neutrinos are Majorana
particles, $c^{}_a \equiv \cos\theta^{}_a$ and $s^{}_a \equiv
\sin\theta^{}_a$, and so on. Current experimental data indicate that
at least two lepton mixing angles are much larger than the Cabibbo
angle $\theta^{}_{\rm C} \simeq 13^\circ$ \cite{PDG}. In
particular, the tri-bimaximal flavor mixing pattern \cite{TB}
\begin{eqnarray}
U^{}_0 = \left( \matrix{ \frac{\sqrt 2}{\sqrt 3} & \frac{1}{\sqrt 3}
& 0 \cr -\frac{1}{\sqrt 6} & \frac{1}{\sqrt 3} & \frac{1}{\sqrt 2}
\cr \frac{1}{\sqrt 6} & -\frac{1}{\sqrt 3} & \frac{1}{\sqrt 2} \cr}
\right) P^{}_\nu
\end{eqnarray}
is quite consistent with the observed values for the solar and
atmospheric neutrino mixing angles and can easily be derived from a
number of neutrino mass models based on discrete flavor
symmetries \cite{Review}. Comparing Eq. (15) with Eq. (16), we
see that they become equivalent to each other if the conditions
\begin{eqnarray}
~~~~ \theta^{}_a \simeq 54.7^\circ \; ,
~~~~ \theta^{}_b = 45^\circ \; , ~~~~ \theta^{}_c = 60^\circ \; ,
~~~~ \phi^{}_l = 0^\circ \;
\end{eqnarray}
are satisfied. A particularly interesting point is that the relation
$\tan\theta^{}_b = \tan\theta^{}_c \cos\theta^{}_a$ exactly holds
and thus the matrix element $U^{}_{e3} = -c^{}_a c^{}_b s^{}_c +
s^{}_b c^{}_c e^{-i\phi^{}_l}$ automatically vanishes as $\phi^{}_l$
approaches zero. This observation, together with the promising
ansatz for the quark flavor mixing discussed in Eqs. (11) and (12),
motivates us to consider the following lepton flavor mixing ansatz:
\begin{eqnarray}
~~~~ U = \left( \matrix{ \frac{1}{2\sqrt{6}} \left( 1 + 3
e^{-i\phi^{}_l} \right) & \frac{1}{\sqrt 3} & -\frac{1}{2\sqrt{2}}
\left( 1 - e^{-i\phi^{}_l} \right) \cr -\frac{1}{\sqrt 6} &
\frac{1}{\sqrt 3} & \frac{1}{\sqrt 2} \cr -\frac{1}{2\sqrt{6}} \left
( 1 - 3 e^{-i\phi^{}_l} \right) & -\frac{1}{\sqrt 3} &
\frac{1}{2\sqrt{2}} \left( 1 + e^{-i\phi^{}_l} \right) \cr} \right)
P^{}_\nu \; ,
\end{eqnarray}
which reproduces the tri-bimaximal flavor mixing
pattern $U^{}_0$ in the $\phi^{}_l \to 0$ limit. In other words, the
generation of nonzero $U^{}_{e3}$ is directly correlated with the
nonzero CP-violating phase $\phi^{}_l$ (or vice versa). Similar to
the case of quark flavor mixing, all three lepton mixing angles are
comparably large in this parametrization. Hence it also assures the
``minimal CP violation" in the lepton sector, although one has
not yet observed CP-violating effects in neutrino oscillations.

One may similarly calculate the Jarlskog invariant of leptonic CP
violation and off-diagonal asymmetries of $U$ in modulus based on
Eqs. (15) and (18). The results are
\begin{eqnarray}
J^{}_l = {\rm Im}\left(U^{}_{e1} U^{}_{\mu 2} U^*_{e2} U^*_{\mu 1}\right)
= {\rm Im}\left(U^{}_{e2} U^{}_{\mu 3} U^*_{e3} U^*_{\mu 2}\right) =
\cdots = c^{}_a s^2_a c^{}_b s^{}_b c^{}_c s^{}_c \sin\phi^{}_l =
\frac{1}{12} \sin\phi^{}_l \; ,
\end{eqnarray}
and
\begin{eqnarray}
\Delta^{l}_{\rm L} \hspace{-0.2cm} & = & \hspace{-0.2cm}
|U^{}_{e2}|^2 - |U^{}_{\mu 1}|^2 = |U^{}_{\mu 3}|^2 - |U^{}_{\tau
2}|^2 = |U^{}_{\tau 1}|^2 - |U^{}_{e3}|^2 = s^2_a \left( s^2_c -
s^2_b \right) = + \frac{1}{6} \; , ~~~
\nonumber \\
\Delta^{l}_{\rm R} \hspace{-0.2cm} & = & \hspace{-0.2cm}
|U^{}_{e2}|^2 - |U^{}_{\mu 3}|^2 =
|U^{}_{\mu 1}|^2 - |U^{}_{\tau 2}|^2 = |U^{}_{\tau 3}|^2 - |U^{}_{e1}|^2 =
s^2_a \left( c^2_b - s^2_c \right) = - \frac{1}{6} \; .
\end{eqnarray}
It becomes obvious that the MNSP matrix $U$ is more asymmetric in
modulus than
the CKM matrix $V$, and CP violation in the lepton sector is likely
to be much larger than that in the quark sector simply because the
lepton flavor mixing angles are not suppressed.

To see why the ansatz proposed in Eq. (18)
is phenomenologically interesting in a
more direct way, let us compare it with the standard parametrization
of the MNSP matrix $U$ \cite{PDG}. In this case the neutrino
mixing angles are predicted to be
\begin{eqnarray}
\sin\theta^{}_{13} \hspace{-0.2cm} & = & \hspace{-0.2cm}
\frac{1}{\sqrt 2} \sin\frac{\phi^{}_l}{2} \; ,
\nonumber \\
\tan\theta^{}_{12} \hspace{-0.2cm} & = & \hspace{-0.2cm}
\frac{1}{\sqrt{2 - 3 \sin^2\theta^{}_{13}}} \; ,
\nonumber \\
\tan\theta^{}_{23} \hspace{-0.2cm} & = & \hspace{-0.2cm}
\frac{1}{\sqrt{1 - 2 \sin^2\theta^{}_{13}}} \; .
\end{eqnarray}
So $\theta^{}_{13} \leq 45^\circ$ must hold for arbitrary values of
$\phi^{}_l$. Given the generous experimental upper bound
$\theta^{}_{13} < \theta^{}_{\rm C}$ \cite{PDG}, the upper limit of
$\phi^{}_l$ turns out to be $\phi^{}_l < 37.1^\circ$. A global
analysis of current neutrino oscillation data seems to favor
$\theta^{}_{13} \simeq 8^\circ$ \cite{Schwetz}, implying $\phi^{}_l
\simeq 22.7^\circ$ together with $\theta^{}_{12} \simeq 35.7^\circ$
and $\theta^{}_{23} \simeq 45.6^\circ$. These results are certainly
consistent with the present experimental data. The resulting value
of the leptonic Jarlskog parameter is $J^{}_l = (\sin\phi^{}_l)/12
\simeq 3.2\%$, which should be large enough to be observed in the
future long-baseline neutrino oscillation experiments. Furthermore,
the CP-violating phase $\delta^{}_l$ in the standard parametrization
is found to be much larger than $\phi^{}_l$ in our ansatz:
\begin{eqnarray}
~~~~ \sin\delta^{}_l = \frac{\sqrt{2} \cos^2\theta^{}_{13}}
{\sqrt{2 - 3 \sin^2\theta^{}_{13}}} \; .
\end{eqnarray}
Therefore, we obtain $\delta^{}_l \simeq 84.4^\circ$ for $\theta^{}_{13}
\simeq 8^\circ$. Note again that $\theta^{}_{13} \leq 45^\circ$ holds,
so Eq. (22) is always valid for the experimentally allowed range of
$\theta^{}_{13}$.

The values of the charged-lepton masses at the electroweak scale
have already been given in Ref. \cite{XZZ}, from which we obtain
$m^{}_e/m^{}_\mu \simeq 4.74 \cdot 10^{-3}$ and $m^{}_\mu/m^{}_\tau
\simeq 5.88 \cdot 10^{-2}$. In view of the neutrino mass-squared
differences extracted from current neutrino oscillation experiments
\cite{Schwetz}, we get $m^{}_2/m^{}_3 \simeq 0.17$ in the $m^{}_1
\simeq 0$ limit for a normal mass hierarchy. A naive conjecture is
therefore
\begin{eqnarray}
\sin\phi^{}_l \simeq \sqrt{\frac{m^{}_2}{m^{}_3}} \; ,
\end{eqnarray}
implying $\phi^{}_l \simeq 24.3^\circ$ and thus $\theta^{}_{13}
\simeq 8.6^\circ$. Since $\theta^{}_a$, $\theta^{}_b$ and
$\theta^{}_c$ are all large, it seems more difficult to link them to
the charged-lepton or neutrino mass ratios.

Finally, it is worth pointing out that one may propose similar
ans$\rm\ddot{a}$tze of lepton flavor mixing based on some other
constant patterns with $U^{}_{e3} =0$. For example, we find that the
mixing angles of the democratic \cite{FX96}, bimaximal \cite{BM},
golden-ratio \cite{GR} and hexagonal \cite{Xing03} mixing patterns
expressed in our present parametrization can also satisfy the
condition $\tan\theta^{}_b = \tan\theta^{}_c \cos\theta^{}_a$, and
thus the matrix element $U^{}_{e3} = -c^{}_a c^{}_b s^{}_c + s^{}_b
c^{}_c e^{-i\phi^{}_l}$ automatically vanishes in the $\phi^{}_l \to
0$ limit. Given such a constant pattern, a lepton flavor mixing
ansatz analogous to the one proposed in Eq. (18) can similarly be
discussed. Its salient feature is therefore the prediction
\begin{eqnarray}
\sin\theta^{}_{13} = |U^{}_{e3}| = 2s^{}_b c^{}_c
\sin\frac{\phi^{}_l}{2} \; ,
\end{eqnarray}
which directly links $\phi^{}_l$ to $\theta^{}_{13}$. Given
$\theta^{}_b =45^\circ$ and $\theta^{}_c =60^\circ$ for the
tri-bimaximal flavor mixing pattern, the first
relation in Eq. (21) can then be reproduced from Eq. (24).

\section{Summary}

We have explored a unique parametrization of fermion flavor mixing
in which the mixing angles are nearly democratic and the (Dirac)
CP-violating phase is minimal. Within such a parametrization of the
CKM matrix $V$ we have shown that all three quark mixing angles are
close to the Cabibbo angle $\theta^{}_{\rm C} \simeq 13^\circ$ while
the CP-violating phase $\phi^{}_{q}$ is only about $1^\circ$. It
also provides a simple description of the structure of $V$, which is
almost symmetric in modulus about its
$V^{}_{ud}$-$V^{}_{cs}$-$V^{}_{tb}$ axis. When the MNSP matrix $U$
is parametrized in the same way, we find that the lepton mixing
angles are comparably large (around $\pi/4$) and the Dirac
CP-violating phase $\phi^{}_l$ is also minimal as compared with its
values in the other eight possible parametrizations. These
interesting observations have motivated us to propose a simple and
testable neutrino mixing ansatz which is equal to the well-known
tri-bimaximal flavor mixing pattern in the $\phi^{}_l \to 0$ limit.
It predicts $\sin\theta^{}_{13} = 1/\sqrt{2} \sin
\left(\phi^{}_l/2\right)$ for reactor antineutrino oscillations, and
its two larger mixing angles are consistent with solar and
atmospheric neutrino oscillations. The Jarlskog invariant of
leptonic CP violation is found to be $J^{}_l =
\left(\sin\phi^{}_l\right)/12$, which can reach a few percent if
$\theta^{}_{13}$ lies in the range $7^\circ \leq \theta^{}_{13} \leq
10^\circ$.

It is worth remarking that the unique parametrization discussed in
this paper provides us with a novel description of the observed
phenomena of quark and lepton flavor mixings. Different from other
possible parametrizations suggesting either a ``geometrical" or a
``maximal" CP-violating phase, it allows us to deal with a
``minimal" one. Although it remains unclear whether such a new point
of view is really useful in our quest for the underlying flavor
dynamics of fermion mass generation and CP violation, we believe
that it can at least help understanding the structure of flavor
mixing in a phenomenologically interesting way.

\vspace{0.5cm}

{\bf Note added}: Soon after this paper appeared in the preprint
archive (arXiv:1203.0496), the Daya Bay Collaboration announced
their first $\overline{\nu}^{}_e \to \overline{\nu}^{}_e$
oscillation result: $\sin^2 2\theta^{}_{13} = 0.092 \pm 0.016({\rm
stat})\pm 0.005({\rm syst})$ (or equivalently, $\theta^{}_{13}
\simeq 8.8^\circ \pm 0.8^\circ$) at the $5.2 \sigma$ level
\cite{DYB}. We find that our expectations, such as $\theta^{}_{13}
\simeq 8.6^\circ$ given below Eq. (23), are in good agreement with
the Daya Bay observation.

\vspace{0.5cm}

{\bf Acknowledgement}: Z.Z.X. would like to thank J.M.G. for warm
hospitality at CP3 of UCL, where this work was done. He is also
grateful to J.W. Mei and H. Zhang for useful discussions. The work
of J.M.G. was supported in part by the Belgian IAP Program BELSPO
P6/11. The work of Z.Z.X. was supported in part by the Ministry of
Science and Technology of China under grant No. 2009CB825207, and in
part by the National Natural Science Foundation of China under Grant
No. 11135009.

\end{document}